\documentclass[fleqn,usenatbib]{mnras}

\usepackage{newtxtext,newtxmath}

\usepackage[T1]{fontenc}

\DeclareRobustCommand{\VAN}[3]{#2}
\let\VANthebibliography\thebibliography
\def\thebibliography{\DeclareRobustCommand{\VAN}[3]{##3}\VANthebibliography}

\usepackage{graphicx}	
\usepackage{amsmath}	
\usepackage{siunitx}
\usepackage{comment}
\usepackage{subfig}

\title[First JWST High Redshift Galaxies]{Discovery and properties of ultra-high redshift galaxies ($9<z<12$) in the JWST ERO SMACS 0723 Field}

\author[N. J. Adams et al.]{
N.J. Adams$^{1}$\thanks{E-mail: nathan.adams@manchester.ac.uk},
C.J. Conselice,$^{1}$
L. Ferreira,$^{2}$
D. Austin,$^{1}$
J.\@ A.\@ A.\@ Trussler,$^{1}$
I. Juodžbalis,$^{1}$ 
S. M. Wilkins,$^{3,4}$
\newauthor
J. Caruana,$^{4,5}$
P. Dayal,$^{6}$
A. Verma,$^{7}$
A. P. Vijayan$^{8,9,3}$\\
$^{1}$Jodrell Bank Centre for Astrophysics, University of Manchester, Oxford Road, Manchester, UK\\
$^{2}$Department of Physics and Astronomy, University of Nottingham, NG7 2RD, UK \\
$^{3}$Astronomy Centre, University of Sussex, Falmer, Brighton BN1 9QH, UK\\
$^{4}$Institute of Space Sciences \& Astronomy, University of Malta, Msida MSD 2080, Malta\\
$^{5}$Department of Physics, University of Malta, Msida MSD 2080, Malta\\
$^{6}$ Kapteyn Astronomical Institute, University of Groningen, P.O. Box 800, 9700 AV Groningen, The Netherlands \\
$^{7}$ Sub-department of Astrophysics, University of Oxford, Denys Wilkinson Building, Keble Road, Oxford, OX1 3RH, UK\\
$^{8}$ Cosmic Dawn Center (DAWN)
$^{9}$ DTU-Space, Technical University of Denmark, Elektrovej 327, DK-2800 Kgs. Lyngby, Denmark
}

\date{Accepted XXX. Received YYY; in original form ZZZ}

\pubyear{2015}

\begin{document}
\label{firstpage}
\pagerange{\pageref{firstpage}--\pageref{lastpage}}
\maketitle

\begin{abstract}
We present a reduction and analysis of the \textit{James Webb Space Telescope} (JWST) SMACS~0723 field using new post-launch calibrations to conduct a search for ultra-high-redshift galaxies ($z > 9$) present within the Epoch of Reionisation. We conduct this search by modelling photometric redshifts in several ways for all sources and by applying conservative magnitude cuts ($m_{\rm F200W} < 28$) to identify strong Lyman breaks greater than 1 magnitude. We find four $z > 9$ candidate galaxies which have not previously been identified, with one object at $z = 11.5$, and another which is possibly a close pair of galaxies. We measure redshifts for candidate galaxies from other studies and find the recovery rate to be only 23 per cent, with many being assigned lower redshift, dusty solutions in our work. Most of our $z > 9$ sample show evidence for Balmer-breaks, or extreme emission lines from H$\beta$ and [OIII], demonstrating that the stellar populations could be advanced in age or very young depending on the cause of the F444W excess. We discuss the resolved structures of these early galaxies and find that the S\'{e}rsic indices reveal a mixture of light concentration levels, but that the sizes of all our systems are exceptionally small ($< 0.5$~kpc). These systems have stellar masses M$_{*} \sim 10^{9.0}$ M$_{\odot}$, with our $z \sim 11.5$ candidate a dwarf galaxy with a stellar mass M$_{*} \sim 10^{7.8}$ -- $10^{8.2}$ M$_{\odot}$.   These candidate ultra high-redshift galaxies are excellent targets for future NIRSpec observations aimed to better understand their physical nature.  
\end{abstract}

\begin{keywords}
galaxies: evolution -- galaxies: formation -- galaxies: high-redshift
\end{keywords}



\section{Introduction}

Following the successful launch of the \textit{James Webb Space Telescope} (JWST) in December of 2021 and the beginning of full scientific operations in July 2022, we have now arrived in a new epoch of extragalactic astronomy. Over the next decade, substantial new datasets probing the near-infrared and mid-infrared wavelength regimes of extragalactic emission across most of the age of the Universe are being made available. As our understanding of not only extragalactic science, but also of this new instrumentation improves, it is likely that this field of study will move rapidly as various survey teams generate various observing strategies and analysis pipelines. 

A key design goal of the \emph{JWST} is to push the redshift frontier and search for galaxies that host the first generation of stars when the Universe was less than 5 per cent of its current age. The aim is to tackle questions regarding the formation and evolution of the first galaxies, black holes, and dark matter halos. The history of high-redshift extragalactic astronomy stretches back to  the work of \citet{Hubble1931}, while today we know of tens of thousands of galaxies beyond a redshift of 4 (corresponding to 10 per cent of the age of the Universe, $\sim1.5$Gyr), and individual galaxies have been found as early as $z \sim 10$ \citep[e.g.][]{Bouwens2011,McLeod2016,Bouwens2016,Oesch2018,Salmon2018,Morishita2018,Stefanon2019,Bowler2020, Harikane2022}.  Yet it remains to be seen how many more galaxies might exist during this epoch of reionisation or even earlier, and what their properties are. Some of the earliest results that emerged from \emph{JWST} observations, show that we are now capable of finding candidate galaxies upwards of $z>13$ \citep{Castellano2022,Naidu2022,Atek2022,Yan2022,Donnan2022}.

\emph{JWST} provides extensive new capabilities in the near-infrared, enabling us to reach depths many magnitudes deeper than was previously possible with the other near- and mid-infrared facilities over the last decade (e.g., \emph{Hubble Space Telescope}, the \emph{Spitzer Space Telescope} and the VISTA telescope). Such an increased depth allows for the Lyman-break and wider rest-frame ultraviolet SED of galaxies with redshifts greater than $z > 9$ to be probed, providing insight into not only the redshift of these systems but also their stellar mass and star formation rates. 

The first releases of data from \emph{JWST} on the 12th and 13th July 2022 included the deep data from the RELICS cluster SMACS J0723.3-732 \citep[SMACS 0723,][]{Coe2019,Leo2022}.  We examine  8.7 square arcminutes of this cluster's outer area, with most of this within the second NIRCam module, with very deep imaging (up to $5\sigma$ AB magnitude 28.45 in F444W) in the near-infrared. This is thus an ideal initial dataset with which to conduct a search for very high redshift galaxies.  Because the depth in SMACS 0723 is comparable or deeper than other early release targets, these data can be used to determine some of the early demographics of distant galaxies. Furthermore, we are able to make some of the first comparisons with theory \citep{dayal2022}.

In this paper we present the initial results of our search for $z > 9$ galaxies within the SMACS 0723 field. We explain our methodology, our completeness calculations, and the basic properties of the galaxies we discovered. We also describe in detail our methodology and how we are able to determine  with some certainty that these systems are at high redshift based on their photometric redshift, colours, and properties of their SEDs. The ultimate understanding of the role of galaxies in the early Universe, including at the epoch of reionisation, will require building up large samples at these redshifts.  Studies such as these are the first step in this process with \emph{JWST}, which will ultimately address fundamental questions about how reionisation occurred and when and how the first galaxies assembled. 
 
The structure of this paper is as follows.  In Section~\ref{sec:data}, we describe the SMACS 0723 observational programme, focusing on the parallel NIRCam observation which we have reprocessed, as well as the data products derived from this new data set. In Section~\ref{sec:method} we describe the selection procedure undertaken to define a robust sample of galaxies with redshifts greater than $z > 9$. We present an initial analysis of the completeness using our procedures and describe the properties of the galaxies we have found. We present a summary of our findings in Section \ref{sec:conclusions}.  Throughout this work, we assume a standard cosmology with $H_0=70$\,km\,s$^{-1}$\,Mpc$^{-1}$, $\Omega_{\rm M}=0.3$ and $\Omega_{\Lambda} = 0.7$ to allow for ease of comparison with other observational studies. All magnitudes listed follow the AB magnitude system \citep{Oke1974,Oke1983}.

\section{Data Reduction and Products} \label{sec:data}

\begin{figure*}
\centering
\includegraphics[width=1.2\columnwidth]{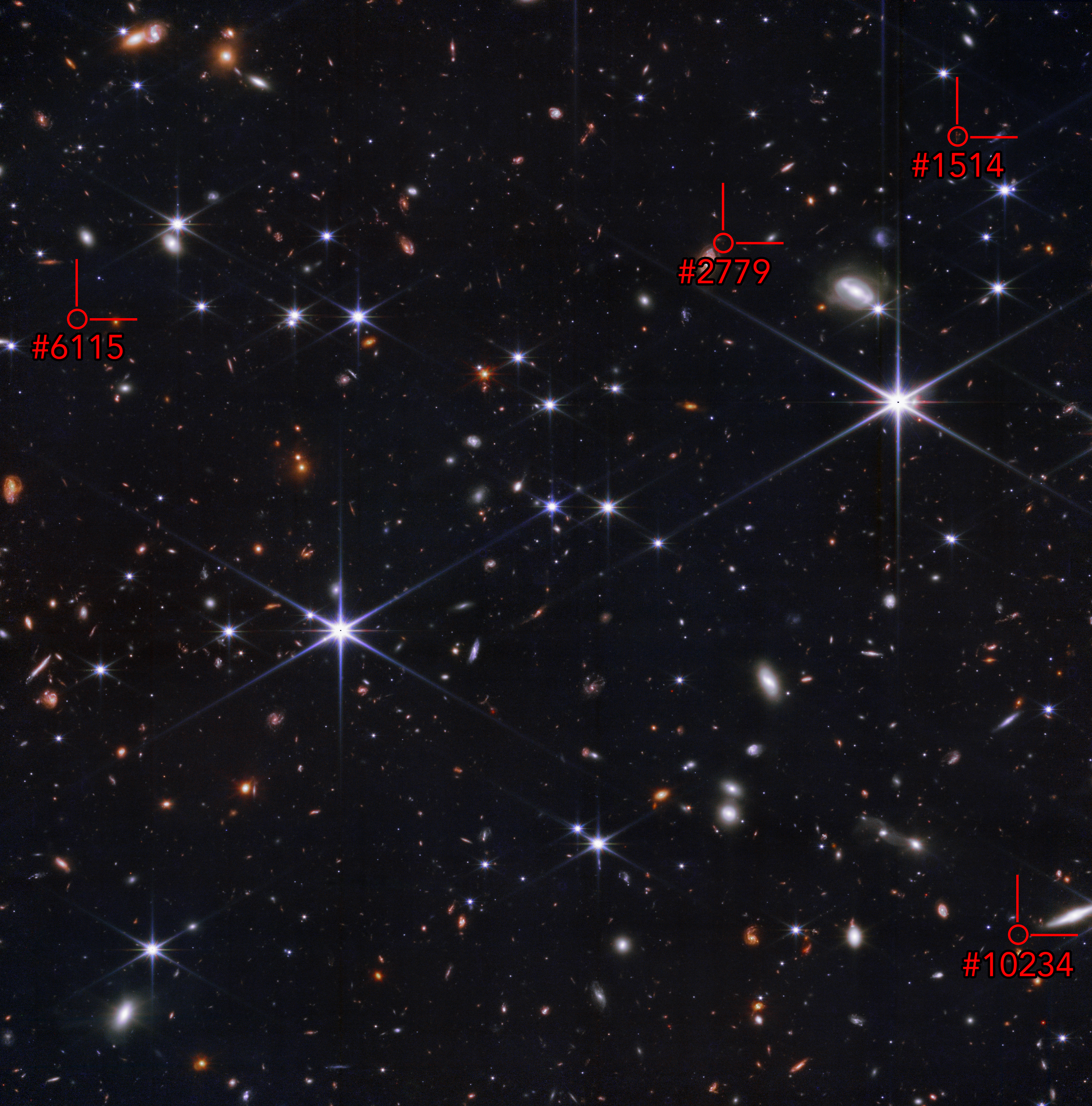}\caption{A colour composite based on our own reduction of NIRCam imaging of the SMACS 0723 parallel field, located approximately 3 arcminutes due South West of the cluster itself. This composite utilises data acquired in 6 bands: F090W, F150W, F200W, F277W, F356W, and F444W, each band being assigned a different colour, sequentially from blue to red. F090W and F150W were mapped to blue hues, F200W and F277W to green, and F356W and F444W to orange and red respectively. The four high redshift candidates that are newly identified in this work are  all located in this field. They are highlighted in red circles alongside their ID numbers from our catalogues. The image is approximately $140\times140$ arcseconds in size and centred on the coordinates RA = $110.6859^{\circ}$, Dec = $73.4815^{\circ}$.}
\label{fig:smacs_0723_parallel_field}
\end{figure*}

The data we use for this analysis originates from the Early Release Observations of SMACS~0723 and include observations taken with the \textit{Near Infrared Camera} \citep[NIRCam;][]{rieke05, rieke08, rieke15}. The images were obtained on June 06, 2022 \citep[PI: Pontoppidan; Program ID 2736,][]{Pontoppidan2022} in the F090W, F150W, and F200W short-wavelength (SW) bands, and F356W, F277W, and F444W long-wavelength (LW) bands. The total integration time for this target is 12.5\,hr.  Figure~\ref{fig:smacs_0723_parallel_field} shows the combined colour image of the second NIRCam module which was not centred on SMACS~0723 that we created from our own reduction.

We reprocess all of the uncalibrated lower-level \emph{JWST} data products for this field following our modified version of the \emph{JWST} official pipeline. This is similar to the process used in \citet{Leo2022} but with minor changes and can be summarised as follows: (1) We use version 1.6.2 of the pipeline with the Calibration Reference Data System (CRDS) version 0942 which was the most up-to-date version at the time of writing. Use of CRDS 0942 is essential for zero point issues we discuss in Section \ref{sec:ZP}  (2) We apply the 1/f noise correction derived by Chris Willott on the resulting level 2 data of the \emph{JWST} pipeline.\footnote{\url{https://github.com/chriswillott/jwst}} (3) We extract the sky subtraction step from stage 3 of the pipeline and run it independently on each NIRCam frame, allowing for quicker assessment of the background subtraction performance and fine-tuning. (4) We align calibrated imaging for each individual exposure to GAIA \citep{GAIADR3} using \texttt{tweakreg}, part of the DrizzlePac python package.\footnote{\url{https://github.com/spacetelescope/drizzlepac}}. (5) We pixel-match the final mosaics with the use of \texttt{astropy reproject}.\footnote{\url{https://reproject.readthedocs.io/en/stable/}} The final resolution of the drizzled images is 0.03 arcseconds/pixel. There is rapid development in the above procedure, and so we anticipate future studies to continue to make refinements to the \emph{JWST} pipeline.

Based on the exposure times of the observations conducted of the SMACS 0723 cluster, an initial estimate of the depths that can be achieved in an aperture was obtained through the use of the \emph{JWST} exposure time calculator. Based on a 0.32 arcsecond circular diameter aperture, the expected depths are F090W = 28.55, F150W = 28.90, F200W = 29.10, F277W = 29.15, F356W = 29.15, and F444W = 28.85. Confirming the data reaches these depths, however, is a challenge. Specifically within this field, there is the lensing cluster itself, taking up close to 50 per cent of the usable area with a high density of sources and intra-cluster light. Additionally, there is a large number of very bright stars (GAIA $m_g<18.5$) inside and on the periphery of the field, creating criss-crossing diffraction spikes that span over a significant portion of the images (much further than can be initially seen from Figure \ref{fig:smacs_0723_parallel_field}). On top of this, there is also the present uncertainty in the behaviour of the telescope's instruments and its reduction pipeline. 

Of particular concern is the handling of flat-fielding, background subtraction, and so-called 1/f noise. We have trialled multiple proposed methodologies for dealing with the above, with mixed results.  Imperfections, such as faint vertical/horizontal banding or gradients across the image, have a negative influence on commonly used methodologies to calculate the $5\sigma$ depth of the image in each photometric band. This includes placing empty circular apertures into the images or using a source subtracted map to examine the RMS of majority sky pixels. We used both of these methods on our images with masking implemented using the segmentation maps, which we derive later on in this paper, as well as manual masking of prominent diffraction spikes and 1/f banding that is still visible after our attempts to correct them. 

Examining the RMS of sky pixels in our unmasked areas, we estimate depths of F090W = 28.2, F150W = 28.2, F200W = 28.4, F277W = 28.9, F356W = 28.9, and F444W = 28.45 in 0.32 arcsecond diameter circular apertures. Compared to the theoretical estimates, it is clear that the pipeline has performed better during the processing of the red module of NIRCam than the blue module. All considered, it is likely that the bluest bands can be pushed deeper with a refined reduction pipeline in the future.

\subsection{Source Photometry and Cataloguing}

\begin{table}
\caption{A list of the primary parameters used in the running of SExtractor on our imaging.}
\centering
\begin{tabular}{c|c}\label{tab:SE}
Parameter         & Value \\ \hline
DETECT\_MINAREA   & 9     \\
DETECT\_THRESH    & 1.8   \\
DEBLEND\_NTHRESH  & 32    \\
DEBLEND\_MINCOUNT & 0.005 \\
BACK\_SIZE        & 64    \\
BACK\_FILTERSIZE  & 3     \\
BACKPHOTO\_TYPE   & LOCAL \\
BACKPHOTO\_THICK  & 24    \\ \hline
\end{tabular}
\end{table}

With the final mosaics of the field completed, we carry out source identification and extraction. For this process, we use the code {\tt SExtractor} \citep{Bertin1996}. We generate multiple different source catalogues for use in calculating the total source counts in each photometric band as a sanity check on our reduction process, with typical detection parameters shown in Table \ref{tab:SE}. For the purpose of this paper, we generate a catalogue which applies forced photometry on the images, using the F444W band as the selection band. This is specifically optimised for the detection of high-redshift galaxies at $z>9$. Output catalogues include basic information about source positions and sizes along with forced aperture photometry. This aperture photometry is calculated within 0.32 arcsecond diameter circular apertures and is corrected with an aperture correction derived from simulated \texttt{WebbPSF} point spread functions for each band used \citep{Perrin2012,Perrin2014}. This diameter was chosen to enclose the central/brightest $70-80$ per cent of the flux of a point source, enabling us to use the highest SNR pixels to calculate galaxy colours while avoiding a reliance on a strong aperture correction.

\subsection{Zero Points \& Pipeline Updates}\label{sec:ZP}

With the release of Calibration Reference Data System (CRDS) version 0942 (operation date 2022-07-29), photometric calibrations from in-flight testing \citep{Rigby2022} were applied to the NIRCam reduction pipeline. This contains a significant shift in the conversion from pixel counts to MJy/sr, the units which the final, reduced images are produced in. The origin of this is primarily the over-performance of the red (long wavelength) NIRCam modules by around 20 per cent. This means that all image reductions conducted using the pre-flight calibrations will observe systematically brighter fluxes for their sources in the red side of the NIRCam instrument. We conduct a comparison of the fluxes measured for objects in the SMACS field to observe how large these systematic shifts are. The results are provided in Table \ref{tab:ZP}.

We observe shifts of up to 0.27 magnitudes between the catalogues generated on the before and after images. These values were validated by conducting the same procedure on the GLASS NIRCam observations \citep{Treu2022} to which consistent offsets were found within 0.02 magnitudes. Offsets of this size can have important ramifications for studies using imaging generated before the post-flight calibrations were fully implemented. It is likely that further updates will come out as more is understood about this new instrumentation. Therefore, we urge further caution when using early JWST products, as this will impact higher level products such as SED fitting and photometric redshifts. The candidates we present in this work are relatively bright ($m_{\rm F200W}<28$) sources that do not push the full sensitivity capabilities of the telescope. As a consequence of this, we find that we recover all of our final candidates in catalogues generated using pre- and post-flight calibrations with systematic shifts to the final redshift within the photo-$z$ errors.

\begin{table}
\caption{The mean magnitude difference between the catalogues generated using imaging that used the jwst pipeline before CRDS 0942 was released and afterwards. The numbers provided are a magnitude correction to be applied to catalogues we generated before the update to match those generated afterwards.}
\centering
\begin{tabular}{l|ll}\label{tab:ZP}
Filter & Module A & Module B \\ \hline
F090W  & -0.147   & -0.174   \\
F150W  & -0.051   & -0.047   \\
F200W  & -0.128   & -0.114   \\
F277W  & +0.067   & +0.223   \\
F356W  & +0.145   & +0.163   \\
F444W  & +0.272   & +0.162  
\end{tabular}
\end{table}

\section{A Robust Sample of Ultra-High Redshift sources} \label{sec:method}

With the imaging and cataloguing completed, we carried out a search for high-redshift sources embedded deep within the epoch of reionisation (at $z>9$). For this first search, we approach this problem in a conservative manner, employing stringent cuts to ensure a pure and robust sample of candidates. It is for this reason that we employ a magnitude cut of 28 in the F200W band within a 0.32 arcsec diameter aperture in our F444W selected catalogue. This equates to  $7.5 \sigma$ based on our initial depth calculations. Such a bright magnitude is used to ensure that Lyman-breaks are correctly identified as such and are not actually Balmer-breaks at lower redshift. Since the F090W filter is of depth 28.2 (providing $2\sigma$ limits of 29.2), if we were to push the limits of the field with a deeper F200W selection, the F090W filter will not be deep enough for a non-detection to rule out a deep Balmer-break or smoother SED indicative of a lower-z galaxy. Following the same logic, we required objects to have a minimum measured magnitude of 28 in the F277W band for objects with $z>12$, enabling only objects with strong breaks within the F150W and F200W bands to be considered when searching for those with $z>12$. Additionally, the SMACS 0723 observations unfortunately lacks the NIRCam F115W filter, that is being used by nearly every other deep field at present, potentially reducing the accuracy of photometric redshifts and efficacy of colour cuts. As our understanding of both the instrumentation aboard \emph{JWST} and the SEDs of ultra-high redshift galaxies improve, it is likely that we will be able to make full use of the depths available in the near future, but for now there is some risk of faint/low-mass $z\sim2-5$ interlopers falling into the selection criteria if stringent cuts are not made. We are confident that the conservative cuts adopted have successfully removed most of this contamination.

\subsection{Spectral Energy Distribution Modelling and Photometric redshifts}

\begin{figure*}

\centering
\vspace{-0.3cm}
\subfloat[ID 1514]{\includegraphics[width=12cm]{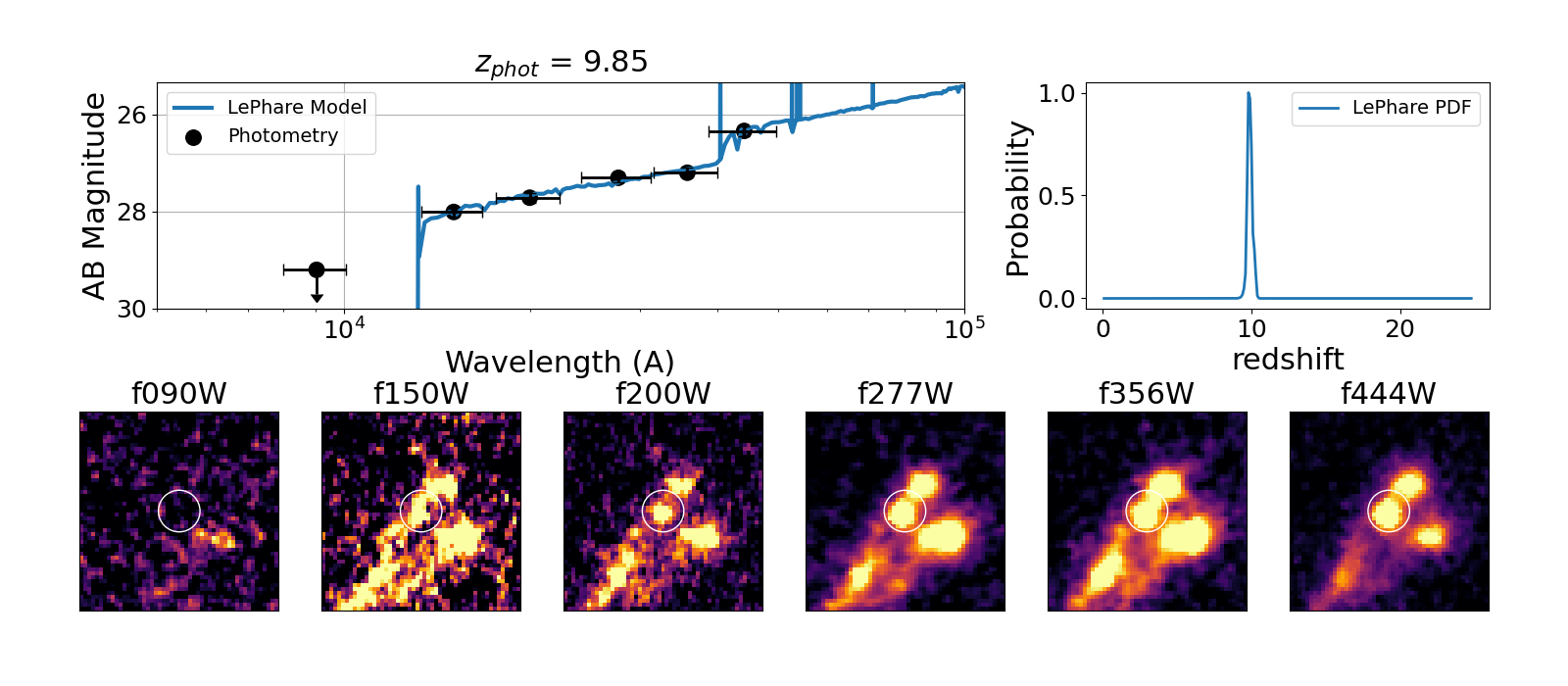}}\hfil 
\vspace{-0.35cm}
\subfloat[ID 2779]{\includegraphics[width=12cm]{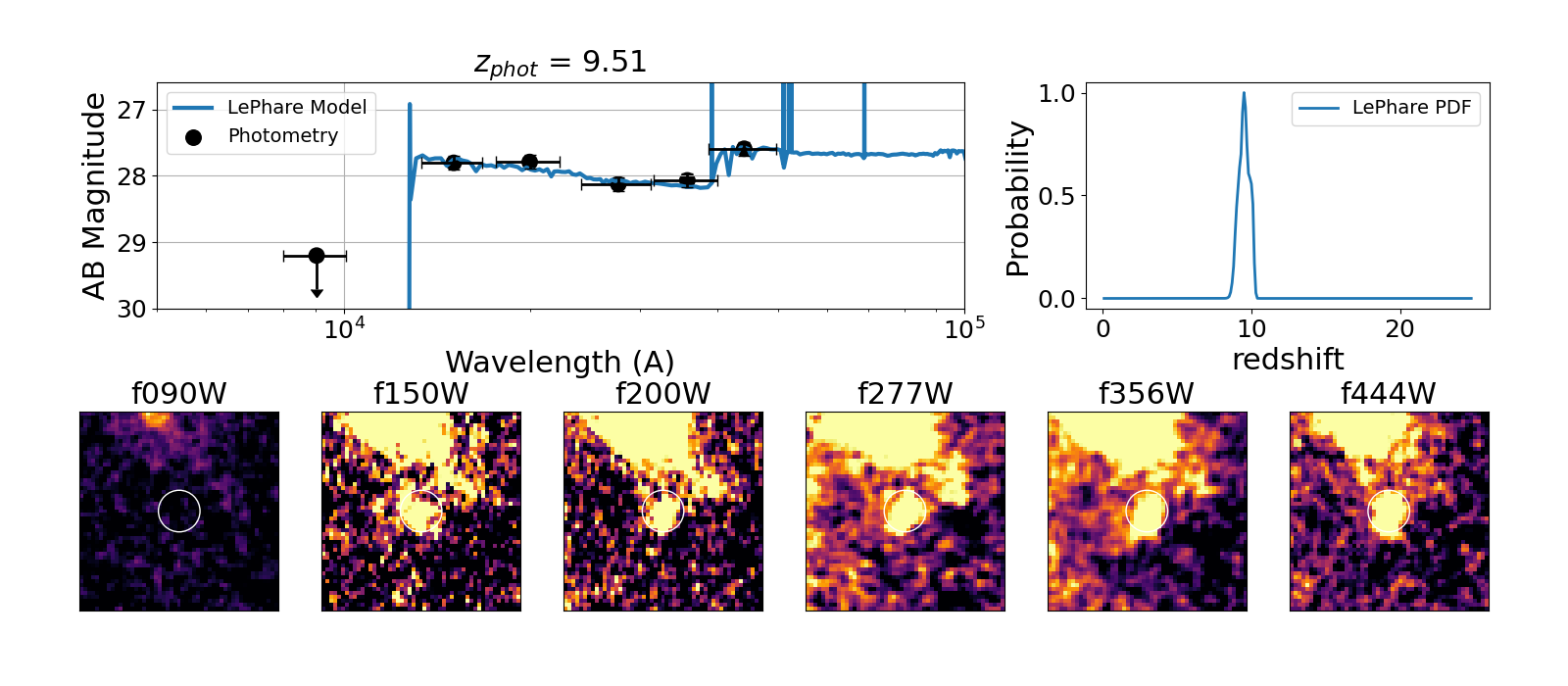}}\hfil
\vspace{-0.35cm}
\subfloat[ID 6115]{\includegraphics[width=12cm]{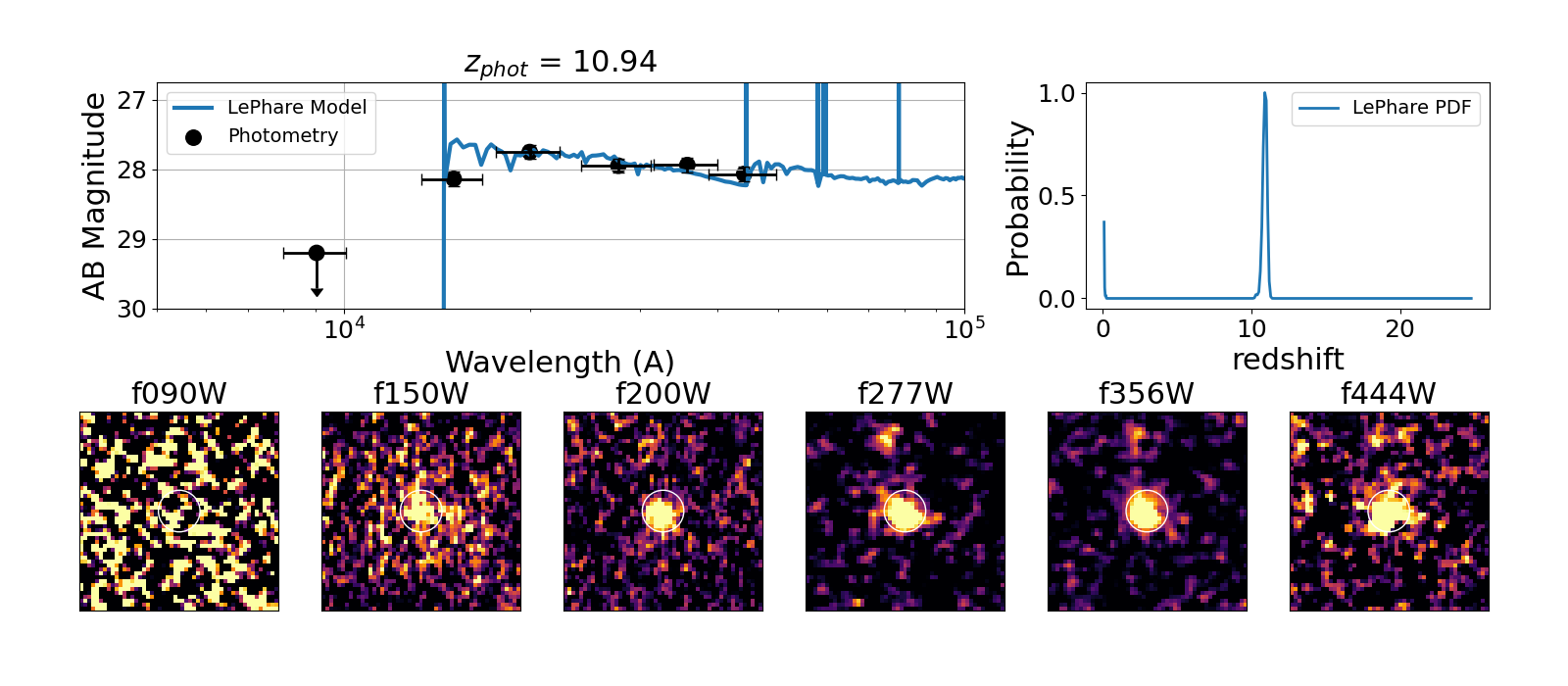}}\hfil
\vspace{-0.35cm}
\subfloat[ID 10234]{\includegraphics[width=12cm]{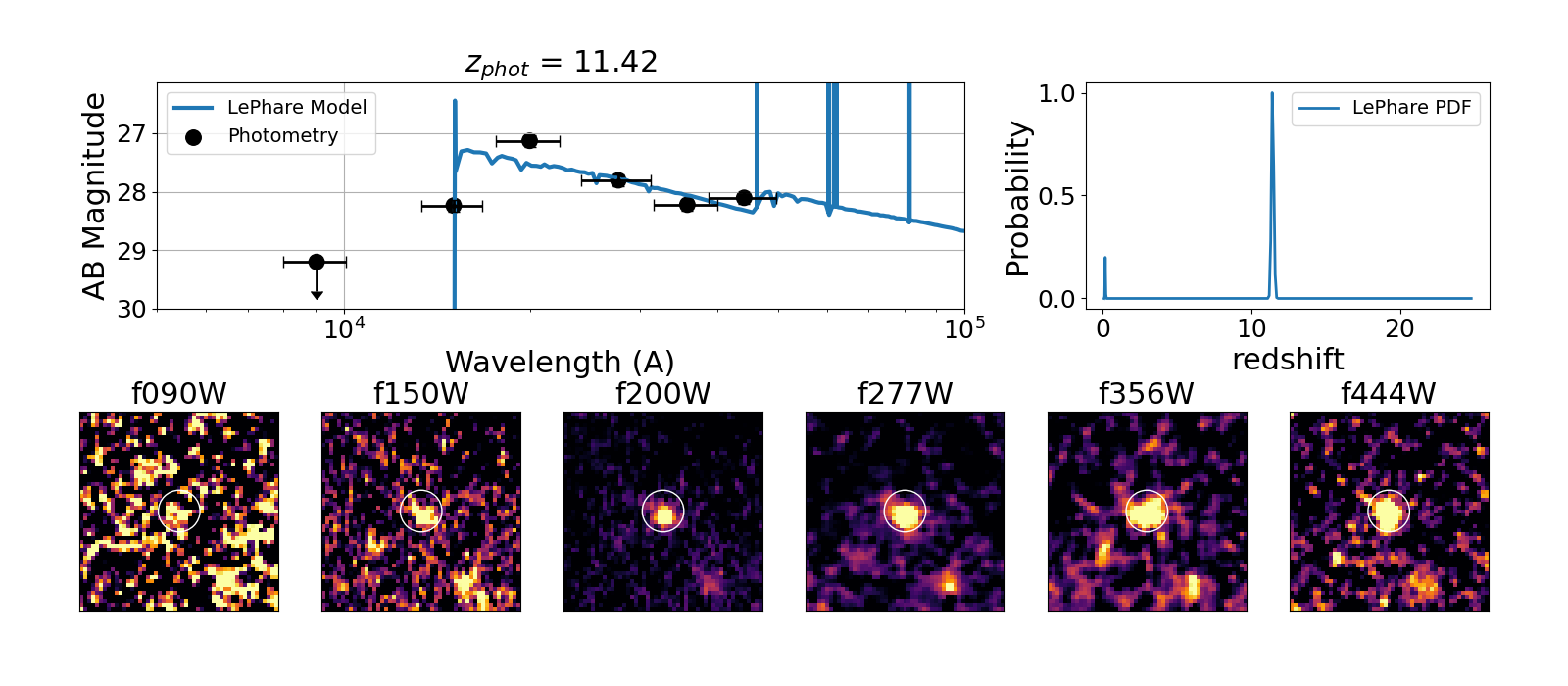}}\hfil
\vspace{-0.05cm}
\caption{Here we present the SEDs for the 4 new candidate $z > 9$ galaxies identified in this study, showing the photometry in the different \emph{JWST} bands as observed with NIRCam.   Overplotted on each panel is the best-fit \textsc{LePhare} model for the SED for these galaxies from which the photometric redshifts are derived. X-axis errors are the filter FWHM.  We also show the PDF of the photometric redshift distribution, and in the bottom panel the images in our six bands. The pixel scale of these images is 0.03 arcseconds per pixel, cut outs are $1.5\times1.5$ arcseconds and are orientated so that north follows the Y-axis.}\label{figure}
\end{figure*}

Our photometric redshifts employ the use of multiple different commonly used codes, each of which makes use of different statistical methodologies. We use \textsc{LePhare} (minimising chi square) with checks from other codes to insure a consistency.

Our first round of SED fitting utilises \textsc{LePhare} \citep{Arnouts1999,Ilbert2006} to conduct an initial measurement of the photometric redshift of the sources in our catalogue with a greater than $7.5\sigma$ detection in the selection band of the catalogue. We set this code up to fit templates from \citet{Bruzual2003} with the \citet{Chabrier2000} IMF, using exponential star formation histories with 10 characteristic timescales between $0.01<\tau<13$~Gyr, and 57 different ages between 0 and~13 Gyr. The redshift range allowed is $0<z<25$, a prior on the absolute UV magnitude is set so that the upper allowed limit is -24 and we apply dust attenuation following \citet{Calzetti2000}. Emission lines are turned on and the treatment for the IGM attenuation derived in \citet{Madau1995} is also applied. The photometry is given a minimum error of 5 per cent.

\begin{figure}
\centering
\includegraphics[width=1\columnwidth]{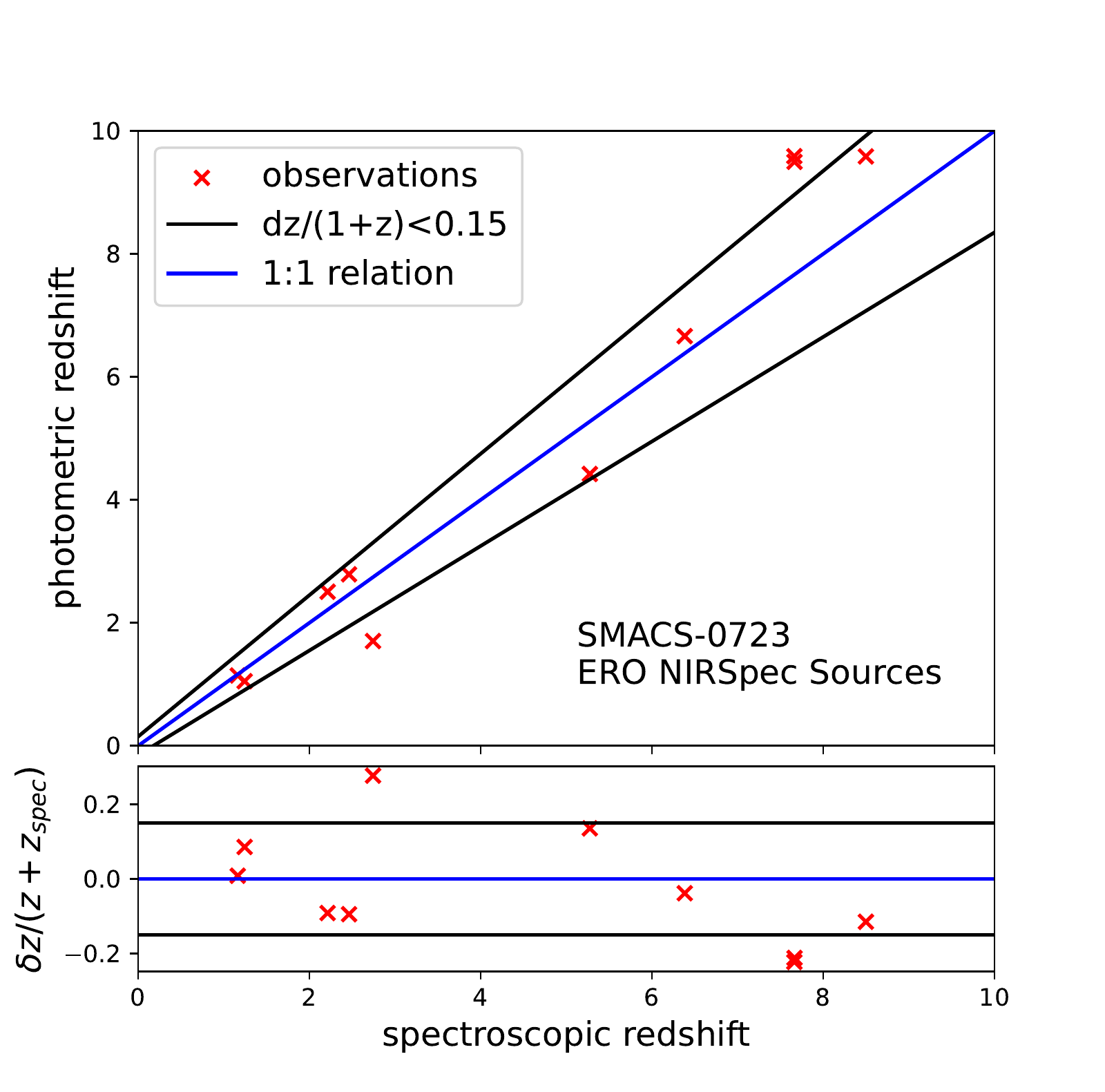}\caption{Initial diagnostics for the performance of the \textsc{LePhare} photometric redshifts.   We compare our photometric redshifts with spectroscopic redshifts for a NIRSpec sample measured by \citet{Carnall2022}.  As can be seen, we find a generally good agreement between these spectroscopic redshifts and our photometric ones, with some scatter.     }
\label{fig:photoz}
\end{figure}

To validate the photometric redshifts, we compare to some of the initial NIRSpec spectroscopic analysis of \citet{Carnall2022}.   We use the 10 spectroscopic redshifts provided by this study, 5 for lower redshift sources and 5 for higher redshift sources. We find that all 10 of these have a cross-match to our F444W selected catalogue. We present the comparison in Figure \ref{fig:photoz}. The agreement between our photo-$z$s and these spec-$z$s is reasonably good, considering the wavelength range available, number of bands in use, and, critically, the lack of the F115W band. A total of 7/10 objects satisfy the typical photometric redshift definition of not being an outlier (Normalised Mean Absolute Deviation \citep[NMAD;][]{hoaglin2000understanding} $|\delta z|/(1+z) < 0.15$). It is worth noting that with the old zero points in the pipeline, only 4/10 objects satisfy the above criteria, showing that the zero point changes have led to a positive outcome for photo-z accuracy. All three objects exhibiting an F090W break are correctly identified as being at high redshifts. However, our redshift estimations are found to be systematically higher. The cause of this is likely the very high H$\beta$ and [OIII] equivalent widths determined in the spectrophotometric analysis conducted in \citet{Carnall2022}. This would inflate the F444W measurement and shift a photometric redshift higher if the lines strengths in the templates are not as high as in reality. There is also the lack of the F115W filter, which \citet{Trussler2022} show significantly improve the photometric redshifts for these spectroscopically confirmed objects.  In either case, our procedure has shown to be effective at identifying F090W break galaxies, though the precise final redshifts are tentative until spectroscopy can be obtained.

For the F444W selected catalogue, the total number of sources meeting an initial cut of $z>9$ and $m_{\rm F200W} < 28$ was 15. These were visually inspected to remove objects that are close to the image edge, that are clearly under the influence of artefacts (such as diffraction spikes) or have a strong secondary peak in redshift. This removed 8 sources which were either contaminated or had strong secondary peaks at $2<z<5$ and were at risk of a Lyman/Balmer-break confusion. Of the 7 remaining, 3 had spectroscopic redshifts previously discussed, leaving us with a final sample of 4 new sources with photometric redshifts greater than 9. The final SEDs of these four new sources are presented in Figure \ref{figure}.

To assess what impact the choice of SED fitting code has on the stellar mass estimations of our candidates, we repeat the above using the {\tt BAGPIPES} code \citep{Carnall2018_Bagpipes}. Here, we fit to galaxies with a constant star formation history and the same aperture corrected photometry used in our \textsc{LePhare} analysis. Priors are left wide and uniform with the exceptions of redshift between $9<z<15$, metallicity set to be between $0.01<Z_\odot<0.5$ and the ionising parameter ($U$) limited to between $-4<\log_{10}(U)<-2$ \citep{Carnall2022}. The results of this process are presented in Table \ref{tab:properties}. We find that half of our objects have a stellar mass consistent within 0.4 dex. Galaxies which disagree the most are those with spectroscopic redshifts, where the different emission line treatments between the codes handle the F356W-F444W colour, and how much if it resembles a Balmer break, differently. Additionally, {\tt BAGPIPES} stellar masses were found to be very sensitive to choice of priors and whether variables were fit in logarithmic or linear space. 

The range of allowed emission line strengths in different template sets and photo-z codes can lead to a wide range of stellar masses (more than a dex) that can be fit to a single object \citep[see e.g. Figure 14 of][]{Endsley2022}. Additionally, there is the possibility that the application of low redshift star formation physics to high redshift sources also biases stellar mass measurements \citep{Steinhardt2022}. Subsequently, stellar mass measurements may need to be taken with a pinch of salt until more is understood about these sources.

Within the early JWST data releases, the work of \citet{Labbe2022} have noted the presence of high redshift galaxies with extremely high stellar masses ($\log(M/M_\odot)\sim10^{11}$) at $7.5<z<10.0$. Such high masses generate tension with the $\Lambda$CDM model of cosmology, as the collapse timescale of gas into dark matter halos, gas density and the efficiency of star formation mean such high stellar masses should be impossible to form \citep{Boylan-Kolchin2022,Lovell2022}. The candidates presented in this work have stellar masses in the regime of $\log(M/M_\odot)\sim10^{8}-10^{10}$, with the highest redshift candidates having the smallest masses. Additionally, the \textsc{LePhare} fitting procedure prevents templates older than the age of the Universe at each redshift from being fit. This means none of our candidates cause significant tension with cosmological models.

\begin{table*}
\caption{A complete list of the coordinates and photometric information for our candidate high-redshift sources. In the first column, we provide the identifying number within our catalogues, the second and third columns detail the right ascension and declination coordinates of the candidate, column 4 provides the redshift as measured in our template fitting process. Column 5 contains a spectroscopic redshift if one was previously available. Columns 6-11 detail the measured aperture-corrected photometry within 0.32 arcsecond diameter circles in the NIRCam imaging.  Table~\ref{tab:properties} lists the physical and morphological properties of these galaxies.}
\hspace{-0.0cm}\begin{tabular}{llllllllllll}\label{tab:Final}
ID & R.A. & Dec. & $z_{\rm phot}$ & $z_{\rm spec}$ & F090W &  F150W & F200W & F277W & F356W  & F444W \\ \hline
1514 & 110.61476 &  -73.47749 &  $9.85^{+0.18}_{-0.12} $  &  -- &   $>29.2$ &  $28.01\pm0.05$ &  $27.72\pm0.05$ &  $27.31\pm0.05$ &  $27.20\pm0.05$ &  $26.34\pm0.05$  \\ 
1696 & 110.83371 &  -73.43448 &  $9.59^{+0.31}_{-0.16}$  & 7.663 &  $>29.2$ &  $26.67\pm0.05$ &  $26.74\pm0.05$ &  $26.95\pm0.05$ &  $26.85\pm0.05$ & $ 26.20\pm0.05$ \\ 
2462 & 110.84429 &  -73.43501 &  $9.50^{+0.17}_{-0.20}$ & 7.665 &   $>29.2$ &  $26.36\pm0.05$ &  $26.27\pm0.05$ &  $26.20\pm0.05$ &  $25.87\pm0.05$ &  $24.98\pm0.05$  \\ 
2779 & 110.64614 &  -73.47591 &  $9.51^{+0.30}_{-0.34}$ & -- &   $>29.2$ &  $27.80\pm0.05$ &  $27.79\pm0.05$ &  $28.12\pm0.05$ &  $28.07\pm0.05$ &  $27.59\pm0.05$ \\ 
6115 & 110.71693 &  -73.46504 &  $10.94^{+0.12}_{-0.15}$  & -- &   $>29.2$ &  $28.14\pm0.06$ &  $27.75\pm0.05$ &  $27.95\pm0.05$ &  $27.94\pm0.05$ &  $28.07\pm0.05$  \\ 
6878 & 110.85914 &  -73.44910 &  $9.59^{+0.65}_{-0.30}$  & 8.498    &  $>29.2$ &  $27.03\pm0.05$ &  $26.96\pm0.05$ &  $27.02\pm0.05$ &  $26.91\pm0.05$ &  $26.32\pm0.05$  \\ 
10234 & 110.66532 &  -73.50180 &  $11.42^{+0.08}_{-0.06}$  & -- &  $>29.2$ &  $28.24\pm0.07$ & $ 27.13\pm0.05$ &  $27.80\pm0.05$ &  $28.22\pm0.05$ &  $28.10\pm0.05$   \\ 
\hline
\end{tabular}
\end{table*}

\subsubsection{Locations of the targets in colour space}

\begin{figure}
\centering
\includegraphics[width=1\columnwidth]{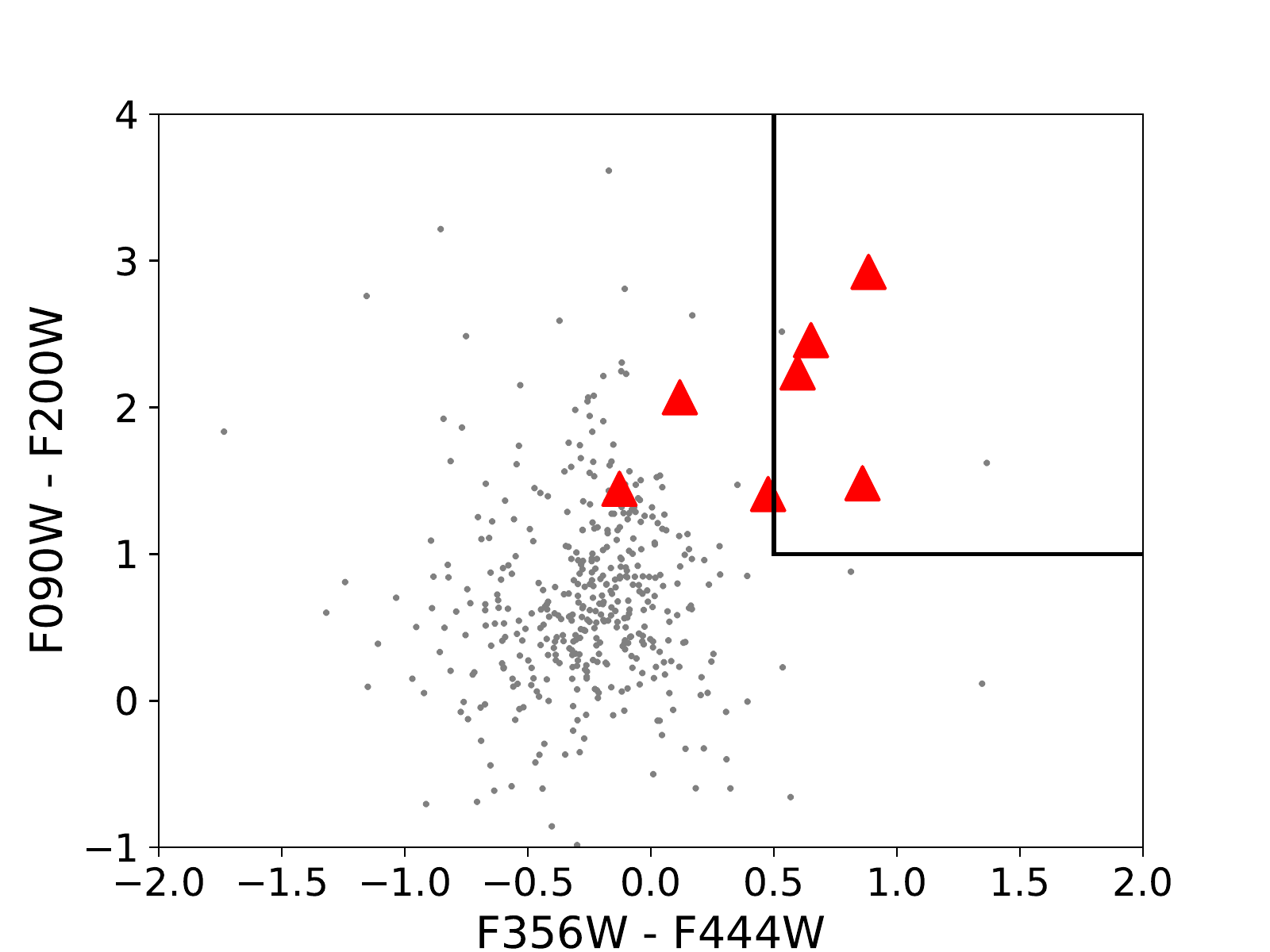}\caption{A colour-colour diagram spanning the Lyman-break (F090W-F200W) and Balmer-break {F356W-F444W} of our high redshift targets and the wider catalogue of objects in the SMACS0723 field. As all of our targets are drop-outs in F090W, the red triangles represent a $2\sigma$ lower limit pointing upwards in the F090W-F200W colour space. The two objects outside of the upper right quadrant are our $z>10.5$ candidates. The grey data points are all other objects with at least a $5\sigma$ F200W detection.}
\label{fig:smacs_colourcolour}
\end{figure}

As a sanity check of our targets, we examine their location in colour space. In particular, we examine the colour space spanning the Lyman- and Balmer-breaks of these objects. The colour-colour diagram comprising these breaks is shown in Figure \ref{fig:smacs_colourcolour} for both our sample and the catalogue as a whole. The upper right quadrant indicates objects with sharp breaks or strong emission lines in SED where Lyman- and Balmer-breaks (or the H$\beta$ and [OIII] lines) would be expected for such high-$z$ targets. We plot on this limits of a break strength of 1.0 mags in the bands spanning the Lyman-break as well as 0.5 mags in the Balmer-break. 

Beyond redshifts of 10.5, however, the Balmer-break has shifted redwards enough such that it is beginning to exit the F444W band. It is thus not surprising that the targets with the weakest Balmer-break are our candidates at $z>10.5$. The redshift for this object is justified on the basis of a 0.4 magnitude or greater drop between the F150W and F200W bands combined with the non-detection in F090W. Objects within this colour region, but not reaching our final sample, have significant contamination, are too faint for our limits, or have a photo-$z<9$.

\subsection{Comparing to Samples from Other Studies}

Within the first couple weeks of the data being released. A number of studies have examined the SMACS 0723 field to search for high-redshift galaxies. Among these are \citet{Atek2022,Yan2022}, two studies that focused on the SMACS 0723 field. Here, we examine the samples found by these studies and examine if there are any consistent objects that appear amongst them. In cases of inconsistency (of which there are many), there is currently little basis to claim who is the more correct on this matter. We instead emphasise that more needs to be done to understand image reductions, sample selections, SED fitting and contaminating populations. However, our independent and careful reductions should be considered within this and ultimately consistency is needed between different studies to verify any sources.

What we find is that the overlap between the redshift candidates found by multiple studies is exceptionally poor. From the 88 candidates in \citet{Yan2022} and 15 in \citet{Atek2022}, only 3 appear in both catalogues and only 2 of those have moderate agreement in photometric redshift ($\delta z < 2$). Currently there are no public catalogues for the entire field made by these teams, but we can make comparisons to the SED fits conducted in this study to begin to investigate why this overlap is so poor. To conduct this comparison, we cross match the catalogues of objects identified in these two studies with our own within a tolerance of 0.5 arcseconds in order to account for different WCS calibrations that have been undertaken. We also test whether there are nearby galaxies with high redshifts in our catalogs that we may have missed. The results of this procedure are presented in Table \ref{tab:compare}.

We begin our comparisons with the sample of objects presented in \citet{Atek2022}. We find a cross match for all 15 candidates with an average separation of 0.2 arcseconds. We find that we have agreement on the photometric redshifts (within 15 per cent) for a total of 6 objects out of the 15. These include the three sources with spectroscopic redshifts at 7.5 and 8.5, meaning only three new sources appear in both of our catalogues. These three are not in the final sample we present here because they fall below our conservative magnitude 28 cuts. They are the objects SMACS-z11e, SMACS-z12b and SMACS-z16b (following \citet{Atek2022} naming). These objects have redshifts 11.52, 12.03 and 15.7 in their original catalogue and we find redshifts of 11.10, 12.09 and 15.39 for these sources. For the other sources reported by \citet{Atek2022} with $z>9$, we find they have solutions with $z<7$ in our analysis. 

Investigating the detailed reasons for this will require a dedicated publication, however, we focus on two of the brighter sources and one ultra-high redshift source to investigate reasons for this significant disagreement. These are SMACS-z10f, SMACS-z11b and SMACS-z16a. For these sources, we obtain photometric redshifts of $z=$ 2.25, 6.94 and 2.96 respectively. For SMACS-z10f, We find a $2\sigma$ detection in F090W and a small excess in F200W, leading to a low-z solution with a balmer break between F150W and F090W alongside bright Balmer series emission lines in F150W and F200W bands. For SMACS-z11b, we also find the source has just under a a $2\sigma$ detection in F090W ($m_{\rm F090W}=29.33$) and has a flux within the F150W band consistent with the measured flux of the F200W band, this is subsequently incompatible with a z=11.7 solution as the Lyman break would be located within the F150W bandpass, and cause a notable reduction in the measured flux. Similarly, we find a substantial detection in the F150W band of magnitude 28.65 for the high-$z$ candidate object SMACS-z16a (raw measurement in 0.32\arcsec aperture, uncorrected for aperture loss). This is consistent with our F200W measurement and the F200W measurement made in \citet{Atek2022}, indicating the break observed between F277W and F200W may be a Balmer one. Examining the F150W imaging by eye shows a faint source is present at the expected coordinates at just under $3\sigma$, while cutouts presented in \citet{Atek2022} indicate no source is present.  This is very likely due to the different reductions in that work and ours.  The source is also in close proximity to diffraction spikes, large foreground objects and near the upper edges of the images. Therefore, any manner of image processing deviations between our works could be the cause of this difference. On the subject of image processing, we find that with the old zero points, candidate SMACS-z11a is given a high redshift solution of $10.29^{+0.22}_{-0.15}$ with the old calibrations, but with the new, post-flight calibrated zero points we obtained a lower redshift solution. This indicates that these candidates can be very sensitive to the processing pipeline.

\begin{table*}
\caption{The results of our comparisons to the photometric redshifts derived in the studies \citet{Atek2022} in the left columns and \citet{Yan2022} in the right columns. We provide the original name assigned to the object and its original redshift. We present alongside the redshift solution found in our study, which we note has processed the images and conducted the SED fitting procedure differently. We display all 15 candidates from \citet{Atek2022} and the 16 matched candidates from \citet{Yan2022} listed as being brighter than magnitude 28 in F444W. The top three \citet{Atek2022} candidates are the three spectroscopically confirmed objects at $z>7$. }
\centering
\begin{tabular}{lll|lll}\label{tab:compare}

Original Name                    & Original $z$   & Our $z$                                & Original Name & Original $z$ & Our $z$             \\[0.03cm] \hline
SMACS-z10a                      & $9.92\pm0.09$  & $9.59^{+0.65}_{-0.30}$                       & F150DB-C-4   & 10.4         & $4.00^{+0.08}_{-0.08}$  \\[0.03cm]
SMACS-z10b                      & $9.79\pm0.2$   & $9.50^{+0.17}_{-0.20}$                       & F200DA-033   & 6.4          & $5.01^{+0.02}_{-0.15}$  \\[0.03cm]
SMACS-z10c                      & $9.94\pm0.1$   & $9.59^{+0.31}_{-0.16}$                       & F150DB-090   & 11.4         & $3.16^{+0.09}_{-0.08}$  \\[0.03cm]
SMACS-z10d                      & $9.98\pm7.97$  & $2.31^{+0.06}_{-0.27}$                        & F150DA-063   & 7.4          & $6.94^{+0.07}_{-0.08}$   \\[0.03cm]
SMACS-z10e                      & $10.44\pm8.34$ & $1.38^{+1.37}_{-0.24}$                       & F150DA-057   & 11.4         & $3.82^{+0.06}_{-0.03}$ \\[0.03cm]
SMACS-z10f                      & $10.47\pm0.47$ & $2.25^{+0.08}_{-0.21}$                       & F150DB-075   & 11.4         & $0.04^{+0.01}_{-0.01}$  \\[0.03cm]
SMACS-z11a                      & $10.75\pm0.28$ & $1.73^{+0.18}_{-0.04}$                       & F150DB-021   & 11.8         & $2.06^{+0.47}_{-0.70}$\\[0.03cm]
SMACS-z11b                      & $11.22\pm0.56$ & $6.94^{+0.07}_{-0.07}$                       & F150DA-050   & 13.4         & $3.42^{+0.30}_{-0.20}$  \\[0.03cm]
SMACS-z11c                      & $11.22\pm0.32$ & $3.84^{+0.05}_{-0.04}$                      & F200DB-086   & 15.4         & $3.53^{+10.28}_{-1.84}$  \\[0.03cm]
SMACS-z11d                      & $11.28\pm3.89$ & $2.35^{+0.30}_{-0.67}$                       & F150DB-041   & 16.0         & $3.70^{+0.02}_{-0.59}$   \\[0.03cm]
SMACS-z11e                      & $11.52\pm9.76$ & $11.10^{+0.21}_{-0.34}$                      & F150DA-083   & 11.8         & $8.68^{+16.32}_{-7.74}$   \\[0.03cm]
SMACS-z12a                      & $12.03\pm0.28$ & $0.10^{+2.26}_{-0.02}$                       & F150DA-058   & 13.4         & $0.12^{+0.03}_{-0.03}$  \\[0.03cm]
SMACS-z12b                      & $12.35\pm0.68$ & $12.09^{+0.16}_{-0.18}$                     & F150DA-062   & 11.4         & $1.78^{+0.20}_{-0.08}$   \\
SMACS-z16a                      & $15.97\pm0.37$ & $2.96^{+0.73}_{-0.21}$                      & F200DB-045   & 20.4         & $0.70^{+0.19}_{-0.05}$   \\[0.03cm]
SMACS-z16b                      & $15.7\pm0.7$    & $15.39^{+0.18}_{-0.26}$                    & F150DA-075   & 13.4         & $2.34^{+0.02}_{-0.02}$   \\
                                &                &                                             & F200DA-098   & 19.8         & $5.20^{+6.34}_{-2.05}$ \\[0.03cm] \hline
\end{tabular}
\end{table*}

Comparing to the catalogue generated by \citet{Yan2022}, we find a similar average WCS offset of 0.2 arcseconds, obtaining a total of 49 cross matches to our F444W selected catalogue. Among these sources, none of the three spectroscopically confirmed galaxies at $7.5<z<8.5$ are included. Of high interest are the reported sources at $z\sim20$. The two brightest sources (F200DB-045 and F200DB-098) are found to have highly complex probability distributions, with multiple peaks. F200DB-045 appears fainter than initially reported, with the F444W measurement a whole magnitude fainter at $m_{\rm F444W}= 29.0$ in our 0.32 arcsecond apertures, meaning the depth of the break is not as significant in our analysis. This is possibly due to different background subtraction procedures that have been applied and their impact on the intra-cluster light (ICL), as this object is located within the lensing cluster. A dedicated study to tackle the foreground cluster and its subtraction is therefore required in order to reveal the nature of this object \citep[e.g.][]{Gu2013,Merlin2016,Livermore2017,Shipley2018,Bhatawdekar2019}. The second object, F200DB-098 is found to be located in close proximity to a diffraction spike. Relaxing our absolute UV magnitude prior, to account for potential lensing effects, improves the redshift agreements for only one object, F200DB-098. For this candidate, we instead obtain a redshift solution of 16.9 with a secondary peak at 19.0. However, this object is not located within the cluster, but in the secondary NIRCam chip and so this would require an extremely luminous galaxy for such a time to be a feasible redshift solution.

The majority of the \citet{Yan2022} sources are very faint, a total of 16 cross matched sources are reported as being brighter than magnitude 28 in the F444W band. Out of all of these objects, we find only one of them has a redshift in agreement with our best-fit photometric redshifts within 15 per cent. This is F150DA-063, which has a reported redshift of 7.4 in \citet{Yan2022} and a redshift of 6.9 in our own. For the remaining objects, we either find a significant F090W detection or extremely red slopes across the NIRCam bands, leading to solutions of very dusty galaxies ($E(B-V)>0.6$) at redshifts 2-5.

To summarise the above, we have found that the overlap of candidate sources between different studies of the same field is very poor. We find that there are two primary reasons for this, the first of which is most relevant to comparisons with our catalogues, where we find we reject sources published in other studies due to the more conservative measures that have been taken regarding apparent luminosity and the proximity of candidates to image edges, artefacts or other luminous objects. The second issue is more widely applicable and is due to conflicting redshift solutions between different studies, with one finding $z>7$ when another finds $z<7$. This is often due to Lyman-Balmer confusion or high levels of dust reddening and the sensitivity of these on the final image reduction. Overall, more needs to be understood about the potential contaminating population of lower redshift sources and the most effective way to process the raw JWST data if we are to be able to push the depths now enabled in the infrared to their fullest extent.

\subsection{Modifying The Fitting Procedure and Templates}

To further assess the candidacy of the high redshift objects that we have identified, we experiment with making small modifications to the photometric redshift procedure and template sets. The results of the following are presented in Table \ref{tab:alternatez}. To begin, we trial switching the photometric redshift code to \textsc{EAZY} \citep{Brammer2008} using a similar BC03 template set made for \textsc{EAZY} that also follows the Chabrier IMF. We label these redshifts ``EAZY\_BC03". We find that the output redshifts are in general agreement with those obtained from \textsc{LePhare}, with all sources having photometric redshifts greater than 8.7.

\begin{table*}
\caption{The results of our modifications to the photometric redshift procedures. For each column we display the best fit redshift and the chi square for the best fit in brackets. For columns 2 and 3, we trial switching on and off the emission line procedure within \textsc{LePhare}. For column 3 we switch the SED fitting code from \textsc{LePhare} to \textsc{EAZY}. Since \textsc{EAZY} does not include an emission line treatment, we manually add emission lines to the templates with two strengths in columns 5 and 6. Column 7 fits combinations of all \textsc{EAZY} templates generated, allowing more variable emission line strengths. Column 8 switches the template set to one designed to incorporate higher redshift physics such as hotter star formation regions. The final column shows the results of fitting exclusively young templates from Wilkins et al. In Prep.}
\begin{tabular}{l|llllllll}
ID    & LePhare & LePhare\_NoLines & EAZY\_BC03  & EAZY\_Lines & EAZY\_Extreme & EAZY\_Combined & EAZY\_HOT & EAZY\_Young \\ \hline
1514  & 9.85 (7.04)   &   9.64 (8.60)               & 10.61 (9.69) & 10.67 (7.95)      & 7.39 (9.75)          & 10.72 (7.36)          & 8.32 (40.23)     &   9.84 (10.5)     \\
1696  & 9.59 (0.05)    & 9.32 (0.20)                 & 8.82 (1.18) & 10.04 (1.27)      & 9.83 (4.22)         & 7.29 (0.87)          & 8.83 (3.09)     & 9.51 (1.44)   \\
2462  & 9.50 (2.08)    & 9.42 (0.90)                 & 9.68 (1.66)  & 10.03 (1.57)      & 9.83 (3.66)         & 9.92 (1.56)          & 8.79 (1.85)      & 9.76 (1.76)  \\
2779  & 9.51 (5.51)    &  9.25 (5.80)               & 8.72 (2.03) & 8.86 (2.53)       & 9.80 (7.04)         & 8.00 (1.54)          & 8.87 (7.16)     &    10.29 (1.79)    \\
6115  & 10.94 (11.25)  & 10.82 (10.20)                 & 10.85 (10.75) & 2.35 (6.61)        & 2.48 (10.54)         & 2.37 (4.03)         & 10.83 (15.12)    &    11.12 (11.19)    \\
6878  & 9.59 (2.91)   &  9.41 (3.40)                & 9.12  (4.43) & 9.85 (4.44)       & 9.82 (8.10)         & 9.12 (4.43)          & 9.07 (5.52)     & 10.05 (4.43)   \\
10234 & 11.42 (80.33)  &  11.41 (83.50)                & 11.28 (104.4) & 2.99 (95.6)        & 2.65 (18.07)         & 2.65 (17.36)          & 2.79  (24.00)    &     11.46 (90.80)  \label{tab:alternatez}
\end{tabular}
\end{table*}

Examining the fit SEDs from both \textsc{LePhare} and \textsc{EAZY}, we find that the candidates at $z\sim9$ are fit with Balmer breaks. As has been shown in \citet{Trussler2022} and \citet{Carnall2022}, an F444W excess can be attributed towards extreme emission lines such as high equivalent width (EW) H$\beta$ and O[III] at a slightly lower redshift of $z\sim8$. To test the impact of these emission lines, we experiment with switching off the emission lines treatment in \textsc{LePhare}. Additionally, the BC03 template set within \textsc{EAZY} does not have emission lines. We thus modify these templates with Gaussian emission lines replicating common high EW lines like Ly$\alpha$, H$\alpha$, H$\beta$, OII and O[III] with two strength levels. The first replicates the EW values of a strong emission line template in \textsc{EAZY}'s Flexible Stellar Population Synthesis \citep[FSPS;][]{Conroy2009,Conroy2010} models (EW$\sim60-100\AA$ ), the second increases the EW values by a factor of 10 to examine a case more akin to the measurements made of $z\sim8$ galaxies. We label these template sets ``EAZY\_Lines" and ``EAZY\_Extreme". An additional template set is used which allows for combinations of all of the preceding templates, allowing for more flexible line strengths, and is called ``EAZY\_Combined".

We find that adding our simple Gaussian emission lines to the \textsc{EAZY} templates cause some shifting of the best fit redshifts. Most notably the higher redshift candidates are now fit with a Balmer break at $z\sim2.5-3$ instead of a Lyman break. The cause is that the SEDs are instead fit with a flat spectrum from F444W to F150W and the F200W is boosted by the strong emission lines that were added to the templates. Examining the best-fit SEDs, the addition of a F115W measurement of comparable depth to the rest of the observations should be capable in constraining if the observed breaks for these objects are from the Balmer series instead of the Lyman series.

The above description for additional emissions lines in the \textsc{EAZY} templates is quite simple and still leaves in room for Balmer breaks to be fit to the SEDs. To test this further, we fit the templates from \citet{Steinhardt2022}, which are a specialist set of templates designed to fit higher redshift sources. They include templates with emission lines designed to match current observations of galaxies at z>8 and include a modification to the initial mass function by increasing the temperature of star forming regions to 45K, as would be expected from an earlier universe with a hotter CMB background. The results from this template set are labelled ``EAZY\_HOT". We find that all but the highest redshift sources retain their high redshift solutions. Though those at $z\sim9$ tend to systematically decrease in redshift due to the lack of Balmer breaks in this template set.

As a final test, we explore the use of a new template set that is currently under development (Wilkins et al. in prep)\footnote{Beta versions of Wilkins et al. templates can be found at \url{https://github.com/stephenmwilkins/synthesizer_eazy_templates}, v0.3 are used in this study}. This template set contains the same BC03 with the Chabrier IMF foundation, but has star formation histories that are exclusively young with mixtures of constant star formation and bursts. The primary purpose of this is to generate template galaxies with very blue ultraviolet SED slopes of $\beta\sim-3$. We label the results from this template set ``EAZY\_Young" and find the redshifts obtained with this template set tend to be slightly higher than found with the broader BC03 template set that we used with both \textsc{LePhare} and \textsc{EAZY}.

\subsection{Completeness  Analysis}

We carry out completeness tests on our data using a simple simulation to extract fake sources that are inserted into the images. To conduct this, we insert galaxies with a Sérsic index of 1 and absolute luminosities ranging from -16.5 to -24 into the F200W and F444W images used for our selection. Half light radii of the fake sources are of different sizes, which are designed to match the theoretical predicted sizes of distant galaxies from BlueTides \citep{Bluetides-I, Bluetides-II,Marshall2022}. The simulation was carried out by sweeping the magnitude range in 0.2 mag intervals, placing 1000 galaxies in the SMACS 0723 parallel field for each interval, and then running SExtractor in the same way on this field as we did for our original detection. A size-luminosity relation of $r \propto L^{0.5}$ was assumed throughout \citep{Grazian2012}, with a reference radius of 800~pc at M$_\mathrm{UV} = -21$ and assuming source redshifts of 11.0.  This allows us to determine how well we can retrieve galaxies of different brightness and concentration at the redshifts of our objects. 

In summary, we find we are able to recover more than 50\% of the simulated galaxies at down to apparent magnitudes of 28.8.  This increases for smaller sized galaxies.   This recovery fraction is 90\% for sources with apparent magnitudes of 28 and is one of the reasons we focus our analysis on candidates brighter than this limit.

\subsection{Galaxy Structure}

\begin{figure}
\centering
\includegraphics[width=0.45\textwidth]{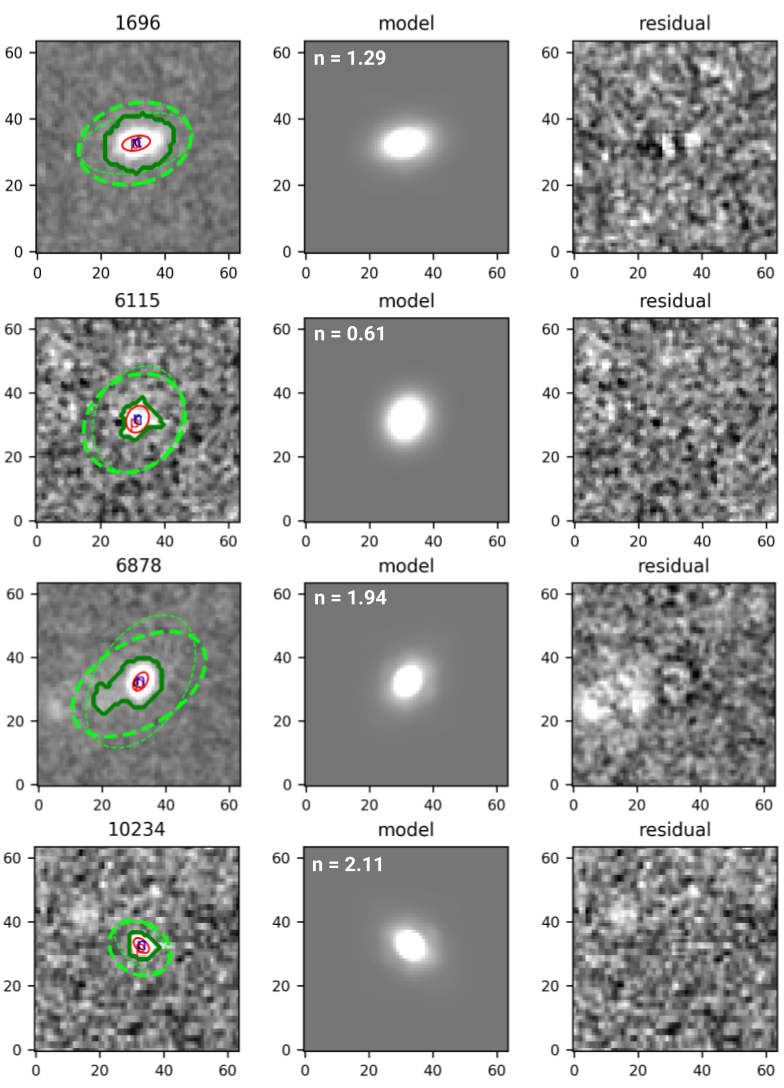}\caption{S\'{e}rsic profile fits for 4 of our high-z candidates. Left panels show the source along with its Petrosian region in dashed light green, the 2D S\'{e}rsic model is shown in the central panel, together with its S\'{e}rsic index ($n$), right panel displays the residuals after source subtraction. Profiles are fitted with the \textsc{Morfometryka} code, as in \citet{Leo2022}.}
\label{fig:sersicfits}
\end{figure}

For the purposes of this work, we limit our structural analysis to the light profiles of our sources through single-component S\'{e}rsic profiles in the F444W band as measured by \text{Morfometryka} \citep{MFMTK}.  We include in this analysis the sources verified as being at $z > 7$ from \citet{Carnall2022}. For our inputs, we used a $64\times64$ pixel stamp as well as a PSF generated with \textsc{webbpsf}. This code measures the luminosity growth curve through aperture photometry in a segmented region based on the Petrosian radius. A 1D S\'{e}rsic fit is performed on the luminosity profile, which in turn is used as inputs to a 2D S\'{e}rsic fit done with the galaxy
and the PSF images.

We list the sizes in terms of the effective radius ($R_\mathrm{e}$) in kpcs as well as the S\'{e}rsic index ($n$), which defines the steepness of the light profile.   These values are listed in in Table~\ref{tab:properties}.  Examples of the fitting procedure, the models, and residuals are shown in Figure \ref{fig:sersicfits}.

\begin{figure}
\centering
\includegraphics[width=0.5\textwidth]{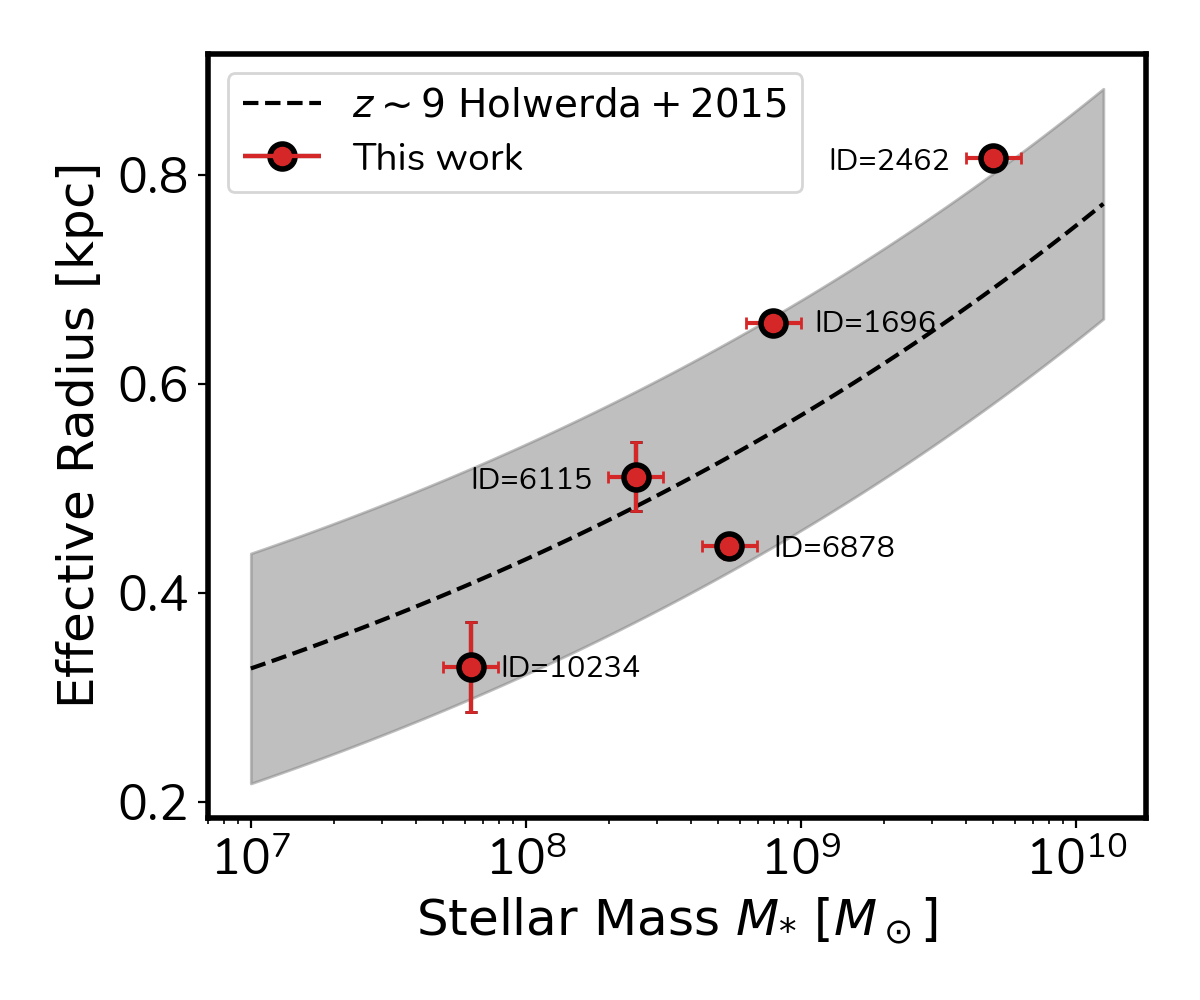}\caption{Effective radius vs stellar mass. Red outline points are the sources reported with structural measurements, dashed black line best fit to size--stellar mass relation from \citet{Holwerda2015} and the shaded region indicates the $1\sigma$ uncertainty on this relation. Masses and redshifts are the base \textsc{LePhare}  stellar mass displayed in Table \ref{tab:properties}.}
\label{fig:sizevsmass}
\end{figure}

Overall, the sources exhibit structures resembling light profiles of disk-like systems, consistent with exponential light profiles. We also find that the galaxies in our sample are very small.  For the most part we find that the effective radii of our systems is $< 0.5$ kpc, with all systems $< 1$ kpc.  Figure \ref{fig:sizevsmass} shows the location of the objects we examine with reference to a previously published $z = 9$ stellar mass vs. effective radius diagram \citep{Holwerda2015}.  Whilst some of our objects are consistent with this relation, one of them appears to be smaller than their stellar mass would imply (ID 6878), this is likely due to the the effects of magnification which we have not corrected for.

\begin{table*}
\caption{The results of the structural fitting process on our list of candidates. Here we display the best fitting photometric redshift from \textsc{LePhare}, unless a spectroscopic redshift is available. The stellar mass is obtained by fixing the redshift to the best redshift and running through \textsc{LePhare}. Alongside we show the size and Sersic index for each galaxy in which a clean fit was possible. Errors on the stellar masses are typically $\pm 0.2$ for the stellar populations models we use. We also present a second set of photometric redshifts and stellar masses derived by running the SED fitting code {\tt BAGPIPES} on the same data. ID 6878 is under the influence of substantial lensing by the foreground cluster, we therefore correct the photometry by a lensing factor of $\mu=5.75$ obtained from the new \citet{Pascale2022} model derived with this new JWST data.}
\centering
\begin{tabular}{ccccccc}
\hline
ID & $z$  & $ \log(M_*/\mathrm{M}_\odot)$  & $n$ & $R_\mathrm{e} \ \mathrm{[kpc]}$ & $z_{\rm Bag}$ & $\log(M_*/\mathrm{M}_\odot)_{\rm Bag}$ \\ \hline
1514 & 9.85  & 9.8  & -& -& $9.94^{+0.12}_{-0.11}$ & 9.7 \\ 
1696 & 7.663 & 8.7 &    $1.28\pm0.05$ & $0.65\pm0.01$ & $7.663$ & 8.1 \\
2462 & 7.665 & 9.5 &   $0.74\pm 0.2$& $0.81\pm0.01$ & $7.665$ & 8.8 \\
2779 & 9.51 & 8.7  & - & - & $9.66^{+0.24}_{-0.31}$ & 8.8 \\ 
6115 & 10.94 & 8.4  & $0.61 \pm {0.14} $ & $0.51 \pm {0.03}$ & $10.81^{+0.13}_{-0.13}$ & 8.0 \\ 
6878 & 8.498  & 8.7   & $1.94\pm0.15$& $0.44\pm0.01$& $8.498$ & 7.8 \\
10234 & 11.42  & 7.8   &  $2.11\pm0.98$ & $0.32\pm0.04$ & $11.25^{+0.08}_{-0.09}$ & 8.2 \\ 
       \hline
       \label{tab:properties}
\end{tabular}
\end{table*}

\subsection{An unusual red structure in the secondary SMACS0723 observation}

\begin{figure*}
\centering
\includegraphics[width=15cm]{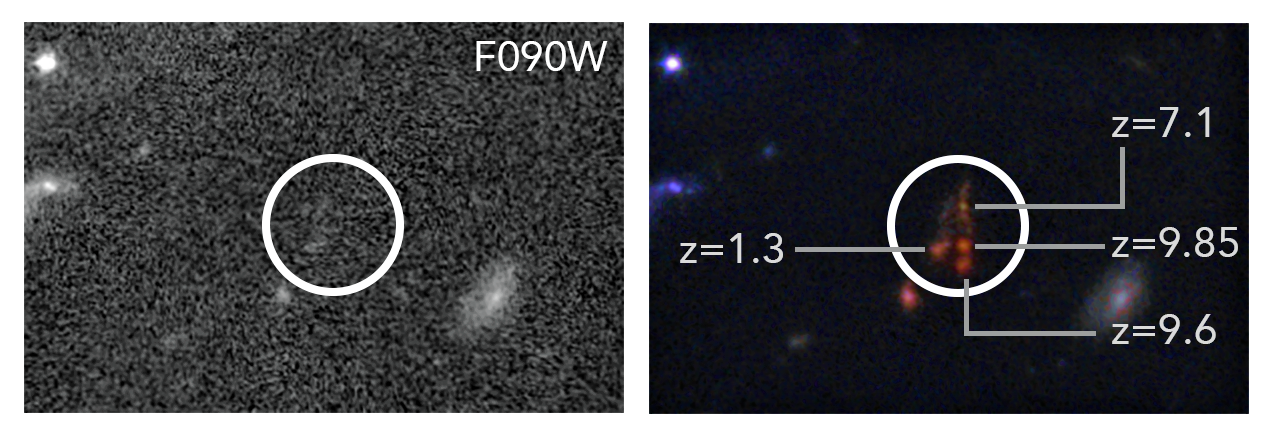}\caption{A zoom into the colour composite presented in Figure \ref{fig:smacs_0723_parallel_field}. The circled object is a large F090W drop out structure composed primarily of four closely neighbouring sources in the `blank' part of the SMACS field. The structure is located at RA=$110.61476^{\circ}$, Dec=$-73.47749^{\circ}$. The left panel shows the F090W image, in which the object(s) drop out, while the right panel shows the colour image created from six filter bands (as described in the caption to Figure~\ref{fig:smacs_0723_parallel_field}). Additionally, the right image shows the four components of the red structure and the best-fit photometric redshifts for each one. The source with a photometric redshift of 9.85 is ID 1514, the neighbouring source with photometric redshift 9.6 falls just below our conservative magnitude 28 cut in F200W. The white circle in both images has a radius of 1 arcsecond.}
\label{fig:smacs_redcutout}
\end{figure*}

Among our final list of high-$z$ candidates is ID 1514 which is located in the NIRCam module that is not centred on the SMACS0723 cluster. In our inspection of our candidates, we find that ID 1514 ($\mathrm{RA}=110.61465$, $\mathrm{DEC}= -73.4774$) is potentially a part of a wider structure located at high redshift. ID 1514 is the bright, compact centre component of the circled system shown in Figure \ref{fig:smacs_redcutout}. Within close physical proximity, the target has three neighbouring objects with similarly very red colours. The object immediately below it has a photometric redshift that is similar to ID 1514 at $z_{\text{phot}}=9.6\pm0.3$; the object immediately above is also a F090W dropout but was provided a redshift of $7.1\pm4.8$ by our template fitting procedure. 

The final object, located to the left, has a weak detection in F090W and a redshift of $1.3\pm0.1$. The exact nature of this structure is presently unknown, but since the majority of this structure drops out in the F090W band, it warrants a mention and further follow-up investigation is planned for the near future.

\subsection{Comparisons to Predictions from Theory}

\begin{figure}
\centering
\includegraphics[width=0.5\textwidth]{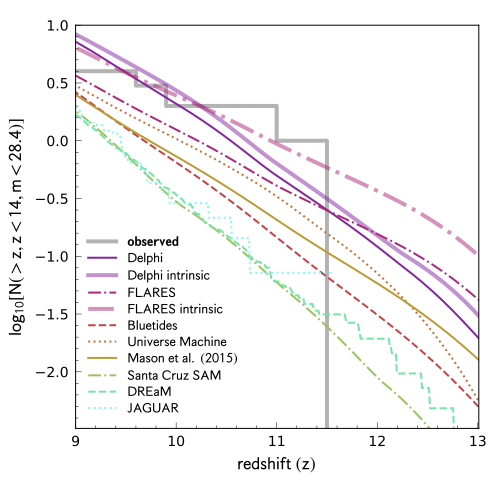}
\vspace{-0.5cm}
\caption{Comparison between different models for how many high redshift galaxies at $z > 9$ we should detect within our probed area of SMACS 0723.  Included here are the predictions from \citep{mason2015}, the semi-analytic model DELPHI \citep[][]{dayal2014, dayal2022} and the Santa Cruz model \citep{yung2019}, as well as hydrodynamical simulations including Bluetides \citep{Bluetides-I, Bluetides-II} and FLARES \citep{FLARES-I, FLARES-II, FLARES-V}.}
\label{fig:theory}
\end{figure}

Although we have just started to probe this epoch in the Universe's history, we can make basic comparisons to theory in terms of how many galaxies we would expect at these magnitudes and redshifts. We carry out a preliminary comparison to a number of models as shown in Fig. \ref{fig:theory}. These include phenomenological models \citep{mason2015}, semi-analytic models 
including DELPHI \citep[][]{dayal2014, dayal2022} and the Santa Cruz model \citep{yung2019} and hydrodynamical simulations including Bluetides \citep{Bluetides-I, Bluetides-II} and FLARES \citep{FLARES-I, FLARES-II, FLARES-V}.   All of these models are shown in Fig. \ref{fig:theory} in terms of the cumulative number of systems at $z > 9$.  What can be seen in all models, and in our data, is that we have fewer systems at higher redshifts given our magnitude limit.  

For a magnitude limit of 28.4 in the NIRCAM F277W filter and an unmasked area of 8.65 square arcseconds, these models predict 2--10 objects at $z \gtrsim 9$ which is in agreement with the 4 sources presented in this work. As expected, for this magnitude limit, the numbers fall off with increasing redshift, with anywhere between 0.03 and 0.5 such objects expected at $z \gtrsim 11.5$; this is somewhat lower than the 1 such source in this work. Finally, we note that the highest amplitudes of the number counts are predicted by DELPHI and FLARES. The DELPHI model being an upper limit is perhaps not surprising since it has been base-lined against the most recent dust estimates from the REBELS (Reionization Era Bright Emission Line Survey) ALMA large program \citep{bouwens2022}. As shown in previous works \citep{dayal2022}, a number of REBELS sources show exceedingly high dust-to-stellar mass ratios leading to a significant attenuation of the UV light; this requires high SFR to be able to reproduce high-$z$ observables. While our number counts are very small and still have large counting and cosmic variance uncertainties (in addition to volume uncertainties due to unaccounted for lensing), this attempt shows that we are now able to compare at these higher redshifts where differences between models can be observed and tested.

\section{Conclusions}\label{sec:conclusions}

We have carried out one of the first analyses of the very early Universe using the Early Release Observations (ERO) data from \emph{JWST} near the cluster field SMACS~0723.  Within this field, we identify 4 new $z > 9$ galaxy candidates based on photometric redshifts and colour cuts, with one object at $z = 11.5$ and another object potentially a part of a more complex, multi-component system at $z\sim9$. The number of galaxies broadly match theoretical predictions for the volume and depth probed by this ERO data. The region occupied by the foreground galaxy cluster has also not been fully analysed, meaning there is further opportunity to find extremely high redshift galaxies that have been strongly lensed in this field.

We find that these systems are robust candidates due to the fact that we find both a Lyman-break and one of a Balmer-break or strong H$\beta$ and O[III] emission; are present within the SEDs of the spectrum of these galaxies. Follow-up spectroscopy will be required to both confirm the redshifts of these galaxies and probe the stellar population. The possible degeneracy in Balmer-breaks and extreme line emission mean that understanding if the stellar population is extremely young or old for the time is difficult to achieve with purely photometric means at this time. By trialling different template sets with different emission line prescriptions, some sources do obtain redshift solutions of $z<9$. However, the inclusion of F115W, which is present in most other deep field programmes planned with JWST, will alleviate this uncertainty due to the improved constraints on the Lyman break location between 1.0-1.5 $\mu$m. Ultimately, more spectroscopic redshifts will be required in order to better investigate the performance of different photo-z template sets and to begin the process of understanding why these varied photo-z estimates are found.

We also find that these early galaxies can have their morphologies well fit by simple parametric fitting and have S\'{e}rsic indices which are around $n = 1$, albeit with large scatter and errors (e.g. our highest redshift source has $n=2.11\pm0.98$), a possible indication of an early structural evolution for galaxies.

Our findings point the way towards the future when large samples of high-$z$ galaxies, such as these and at even higher redshifts, become available from larger area and deeper \emph{JWST} data that is on the horizon. Using these samples we will be able to reconstruct the history of reionisation and the earliest formation of galaxies.

\section*{Acknowledgements}

We thank Anthony Holloway, Sotirios Sanidas and Phil Perry for critical and timely help with computer infrastructure that made this work possible. We also thank the anonymous referee for their constructive comments and discussion.
We acknowledge support from the ERC Advanced Investigator Grant EPOCHS (788113), as well as a studentship from STFC.  This work is based on observations made with the NASA/ESA \textit{Hubble Space Telescope} (HST) and NASA/ESA/CSA \textit{James Webb Space Telescope} (JWST) obtained from the \texttt{Mikulski Archive for Space Telescopes} (\texttt{MAST}) at the \textit{Space Telescope Science Institute} (STScI), which is operated by the Association of Universities for Research in Astronomy, Inc., under NASA contract NAS 5-03127 for JWST, and NAS 5–26555 for HST. These observations are associated with program 14096 for HST, and 2736 for JWST. LF acknowledges financial support from Coordenação de Aperfeiçoamento de Pessoal de Nível Superior - Brazil (CAPES) in the form of a PhD studentship. PD acknowledges support from the NWO grant 016.VIDI.189.162 (``ODIN") and the European Commission's and University of Groningen's CO-FUND Rosalind Franklin program. The Cosmic Dawn Center (DAWN) is funded by the Danish National Research Foundation under grant No. 140.

\section*{Data Availability}

Presently the field is moving very fast. The authors intend to publish a complete set of reduced imaging as well as catalogues of photometric redshifts when the reduction procedure has been completely finalised and an initial set of projects conducted. Those requiring catalogues of the sample discussed in this work are welcome to contact the authors directly for the full set of outputs generated for those objects.



\bibliographystyle{mnras}
\bibliography{mnras_template.bib} 








\bsp	
\label{lastpage}
\end{document}